\documentclass[
 twocolumn,
superscriptaddress,
 amsmath,amssymb,
 aps,
prb,
raggedfooter,
]{revtex4-2}

\usepackage{graphicx}
\usepackage{dcolumn}
\usepackage{bm}
\usepackage[hidelinks]{hyperref}
\usepackage{tabularx}
\usepackage{flushend}
\usepackage{amsmath}
\usepackage{gensymb}

\hypersetup{
  colorlinks   = true, 
  urlcolor     = blue, 
  linkcolor    = blue, 
  citecolor   = red 
}

\begin{document}

\preprint{APS/123-QED}

\title{Persistent Photo-Enhancement of Molecular Orbital Excitations in the Kondo Insulator FeSb$_2$}

\author{Ke-Jun Xu}
\thanks{equal contribution}
\affiliation{%
Material Sciences Division, Lawrence Berkeley National Laboratory, Berkeley, California 94720, USA
}
\affiliation{%
Department of Physics, University of California, Berkeley, California 94720, USA
}

\author{Nathan Giles-Donovan}
\thanks{equal contribution}
\affiliation{%
Department of Physics, University of California, Berkeley, California 94720, USA
}
\affiliation{%
Material Sciences Division, Lawrence Berkeley National Laboratory, Berkeley, California 94720, USA
}

\author{Paulina Majchrzak}
\thanks{equal contribution}
\affiliation{%
Stanford Institute for Materials and Energy Sciences, SLAC National Accelerator Laboratory, 2575 Sand Hill Road, Menlo Park, CA 94025, USA
}
\affiliation{%
Geballe Laboratory for Advanced Materials, Stanford University, Stanford, California 94305, USA 
}

\author{Abhi Suthar}
\affiliation{%
Max-Planck-Institute for Solid State Research, Heisenbergstraße 1, 70569 Stuttgart, Germany
}

\author{Matthias Hepting}
\affiliation{%
Max-Planck-Institute for Solid State Research, Heisenbergstraße 1, 70569 Stuttgart, Germany
}

\author{Marco Trento}
\affiliation{%
Dipartimento di Fisica, Politecnico di Milano, piazza Leonardo da Vinci 32, I-20133 Milano, Italy
}

\author{Giacomo Ghiringhelli}
\affiliation{%
Dipartimento di Fisica, Politecnico di Milano, piazza Leonardo da Vinci 32, I-20133 Milano, Italy
}
\affiliation{%
CNR-SPIN, Dipartimento di Fisica, Politecnico di Milano, I-20133 Milano, Italy
}

\author{Elizabeth Skoropata}
\affiliation{%
PSI Center for Photon Science, Paul Scherrer Institute, Villigen, 5232, Switzerland
}

\author{Hiroki Ueda}
\affiliation{%
PSI Center for Photon Science, Paul Scherrer Institute, Villigen, 5232, Switzerland
}

\author{Eugenio Paris}
\affiliation{%
PSI Center for Photon Science, Paul Scherrer Institute, Villigen, 5232, Switzerland
}

\author{Biaolong Liu}
\affiliation{%
PSI Center for Photon Science, Paul Scherrer Institute, Villigen, 5232, Switzerland
}

\author{Elia Razzoli}
\affiliation{%
PSI Center for Photon Science, Paul Scherrer Institute, Villigen, 5232, Switzerland
}

\author{Wei-Sheng Lee}
\affiliation{%
Stanford Institute for Materials and Energy Sciences, SLAC National Accelerator Laboratory, 2575 Sand Hill Road, Menlo Park, CA 94025, USA
}

\author{Thomas P. Devereaux}
\affiliation{%
Department of Materials Science and Engineering, Stanford University, Stanford, California 94305, USA 
}
\affiliation{%
Stanford Institute for Materials and Energy Sciences, SLAC National Accelerator Laboratory, 2575 Sand Hill Road, Menlo Park, CA 94025, USA
}
\affiliation{%
Geballe Laboratory for Advanced Materials, Stanford University, Stanford, California 94305, USA 
}

\author{Zhi-Xun Shen}
\affiliation{%
Stanford Institute for Materials and Energy Sciences, SLAC National Accelerator Laboratory, 2575 Sand Hill Road, Menlo Park, CA 94025, USA
}
\affiliation{%
Geballe Laboratory for Advanced Materials, Stanford University, Stanford, California 94305, USA 
}
\affiliation{%
Department of Applied Physics, Stanford University, Stanford, California 94305, USA
}
\affiliation{%
Department of Physics, Stanford University, Stanford, California 94305, USA
}

\author{Dung-Hai Lee}
\email{dunghai@berkeley.edu}
\affiliation{%
Department of Physics, University of California, Berkeley, California 94720, USA
}
\affiliation{%
Material Sciences Division, Lawrence Berkeley National Laboratory, Berkeley, California 94720, USA
}

\author{Robert J. Birgeneau}
\email{robertjb@berkeley.edu}
\affiliation{%
Department of Physics, University of California, Berkeley, California 94720, USA
}
\affiliation{%
Material Sciences Division, Lawrence Berkeley National Laboratory, Berkeley, California 94720, USA
}

\date{\today}

\begin{abstract}
When insulators are excited by photons with above-gap energies, photo-doped carriers are transiently generated. Typically, orbital excitations weaken and broaden due to increased screening by these transient carriers, followed by a few-ps recovery via phonon thermalization. Here we report the observation of anomalous dynamics of many-body molecular orbital excitations in the narrow-gap Kondo insulator FeSb$_2$. In the pumped X-ray absorption spectra following excitation with 1~$\mu$m photons, we observe a transient transfer of absorption intensity from the delocalized states at the main absorption peak to the localized molecular orbital states at the pre-edge. At the resonance of the low-energy molecular orbital excitations, the pumped resonant inelastic X-ray spectra show an approximately 30\% intensity increase at a pump fluence of 11 mJ/cm$^2$ concomitant with the emergence of an energy gain feature. While the energy gain feature disappears after about 1-2~ps, the increased molecular orbital excitation intensities partially persist for at least 50~ps with no discernible energy shift. These observations require the invocation of a many-body excitation picture beyond single particle population changes.

\end{abstract}

\maketitle


In Kondo insulators, the Fermi level lies within an energy gap that is derived from a many-body hybridization between localized states hosting magnetic moments and delocalized conduction electrons~\cite{aeppli1, tsunetsugu1}. The hybridization is mediated by the Kondo interaction that gives rise to screening of the local moments and a non-magnetic ground state. At low temperatures, the Kondo gap is well-defined and the many-body interacting state is coherent. Above the Kondo crossover temperature, the conduction electrons scatter from the local moments incoherently~\cite{coleman1}. Conventionally, the local states are constructed from 4$f$ orbital states with a small bandwidth, with prototypical examples being the 4$f$ Kondo insulators SmB$_6$~\cite{menth1, luli1}, YbB$_{12}$~\cite{mignot2}, and Ce$_3$Bi$_4$Pt$_3$~\cite{jaime1}. The entangled Kondo insulating state is highly sensitive to tuning parameters due to the many coupled degrees of freedom, and it is possible to induce quantum phase transitions via external tuning parameters such as doping~\cite{gabani1}, magnetic field~\cite{jaime1}, or pressure~\cite{cooley1}.

\begin{figure*}[t]
\includegraphics[width=0.9\textwidth]{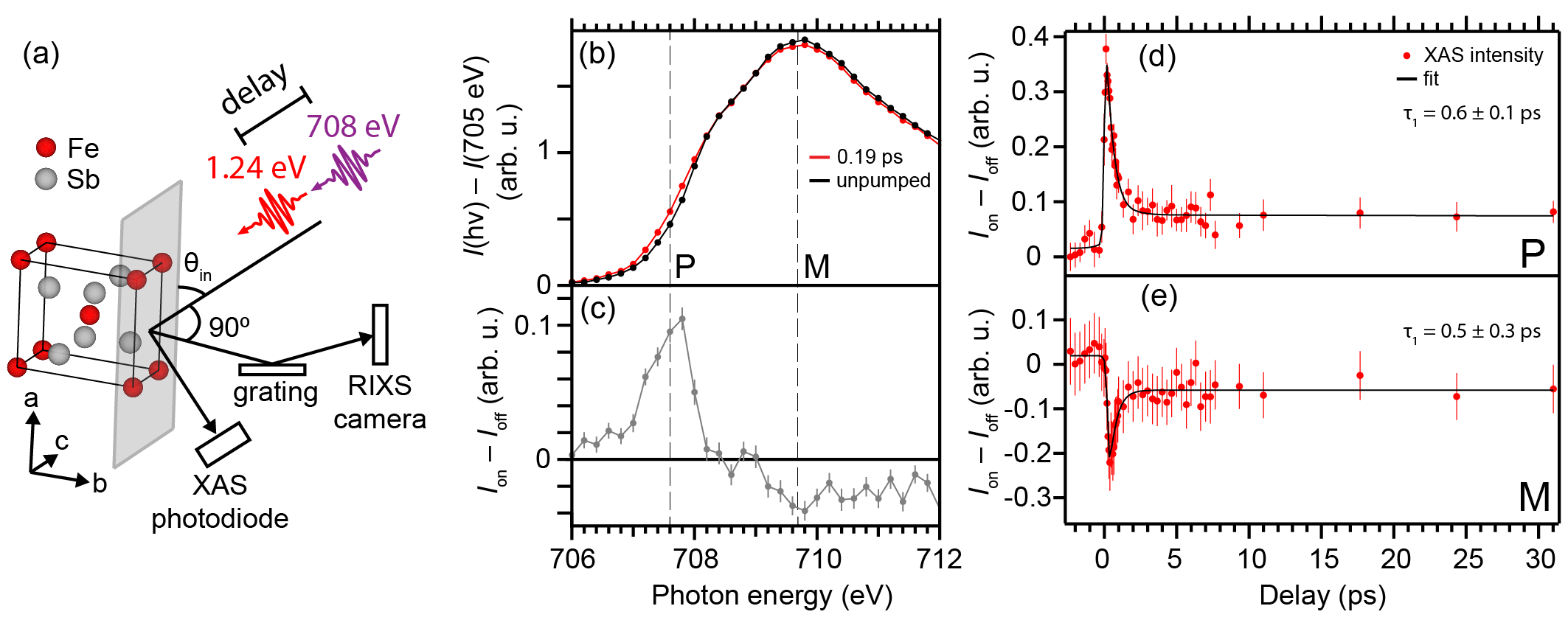}
\caption{Transient X-ray absorption spectroscopy in response to optical excitation. (a) Schematic of the trXAS and trRIXS experimental geometry. The FeSb$_2$ single crystal was cleaved along the $a$-$c$ surface and placed with the $b$ and $c$ axes within the scattering plane. The incidence angle $\theta_{in}$ is the angle of the sample surface relative to the incident X-ray beam. The spectrometer was fixed at 90$\degree$ with respect to the incident beam. (b) XAS spectra for the static condition (black curve) and 0.19~ps after the arrival of the pump pulse (red curve). A constant value (intensity at 705 eV) has been subtracted from each spectrum. (c) Difference between the pumped and static XAS. Dashed vertical lines in (b) and (c) indicate the pre-edge feaure (P) and the main XAS peak (M). (d,e) Time-delay-dependence of the XAS intensity at the P position (d) and at the M position (e). Red dots are the data points and the black line is a fit comprised of a Gaussian rising edge and two decaying exponential trailing edges. Error bars in (c-e) are the 1-$\sigma$ standard deviation. Error bars in (b) are smaller than the data points. Data in all panels taken at 25~K, incidence angle of 70$\degree$, and optical laser excitation fluence of 30.8 mJ/cm$^2$.}
\end{figure*}

Here we study a correlated insulator FeSb$_2$~\cite{Petrovic2}, which exhibits many physical properties~\cite{Petrovic1,Sun1, sun2, eaton1, eo1, kxu2} similar to the prototypical 4$f$ Kondo insulator SmB$_6$. Recently, we have found spectroscopic evidence that FeSb$_2$ hosts localized molecular orbital states that play the role of local moments, along with more delocalized hybridized states that act as the conduction electrons~\cite{kxu1}. In the XAS spectrum, the main absorption peak at the Fe $L_3$-edge is dominated by fluorescence-like emission from continuum states, which can be captured by band structure calculations and the Bethe-Salpeter equations solved by \textsc{ocean}~\cite{vinson1, vinson2, kxu1}. On the other hand, RIXS results show that the pre-edge shoulder of the XAS spectrum is attributed to Raman-like features consistent with molecular orbital states as found in charge transfer multiplet calculations~\cite{kxu1}. Due to the highly hybridized nature of the molecular orbitals, these orbital excitations are $dp$-$dp^*$ excitations rather than conventional $d$-$d$ excitations observed in 3$d$ transition metal insulators~\cite{ament1, degroot1}. These low energy excitations M1 at $\sim$200 meV and M2 at $\sim$800 meV at the XAS pre-edge have been attributed to spin-like and charge-like molecular orbital excitations, respectively. These molecular orbital states acquire anisotropic dispersions through the many-body Kondo interaction with delocalized bands~\cite{kxu1, rahn1}. The M1 and M2 excitations are thought to be analogous to the spin triplet and charge antibonding excitation in the prototypical 4$f$ Kondo insulator SmB$_6$~\cite{kikoin1}.

Here we investigate the effects of ultrafast optical excitations on the many-body ground state in FeSb$_2$. We perform the trXAS and trRIXS experiments at the FURKA endstation of the SwissFEL facility~\cite{razzoli1}. The experimental geometry is shown in Fig. 1(a), where the $b$ and $c$ axes of the FeSb$_2$ sample are oriented within the scattering plane. Here we opted to perform RIXS measurements with the scattering arm fixed at 90$\degree$ with respect to the incident $\pi$-polarized (E field within the scattering plane) photon beam. This geometry suppresses the elastic peak such that the low energy features can be resolved. The grazing geometry accesses high momenta, where the low energy excitations are more separated in energy. We excite the FeSb$_2$ using 1~$\mu$m photons, which have an energy close to the peak in optical conductivity~\cite{homes1} that is associated with an interband transition.

We first characterize the response of the XAS to 1 $\mu$m excitation. Fig. 1(b) shows the absorption spectrum across the Fe $L_3$ edge for the static unpumped condition (black curve) and 0.19 ps after the pump pulse (red curve). We observe a clear increase in intensity at the pre-edge (P) and a small decrease in intensity at the main peak (M). Here, the P feature corresponds to the low energy Raman-like excitations that are attributed to molecular orbital excitations, while the M feature is dominated by the fluorescence-like emission from hybridized band-like continuum states. The difference between the pumped XAS and static XAS is shown in panel (c), where a distinct intensity gain can be observed around P at 707.6 eV. We also observe a broad, nearly constant intensity decrease at incident energies above about 709.5 eV, which suggests that the fluorescence features in this energy range transfer absorption spectral weight to the Raman-like molecular orbital states at P. 

\begin{figure}[t]
\includegraphics[width=0.45\textwidth]{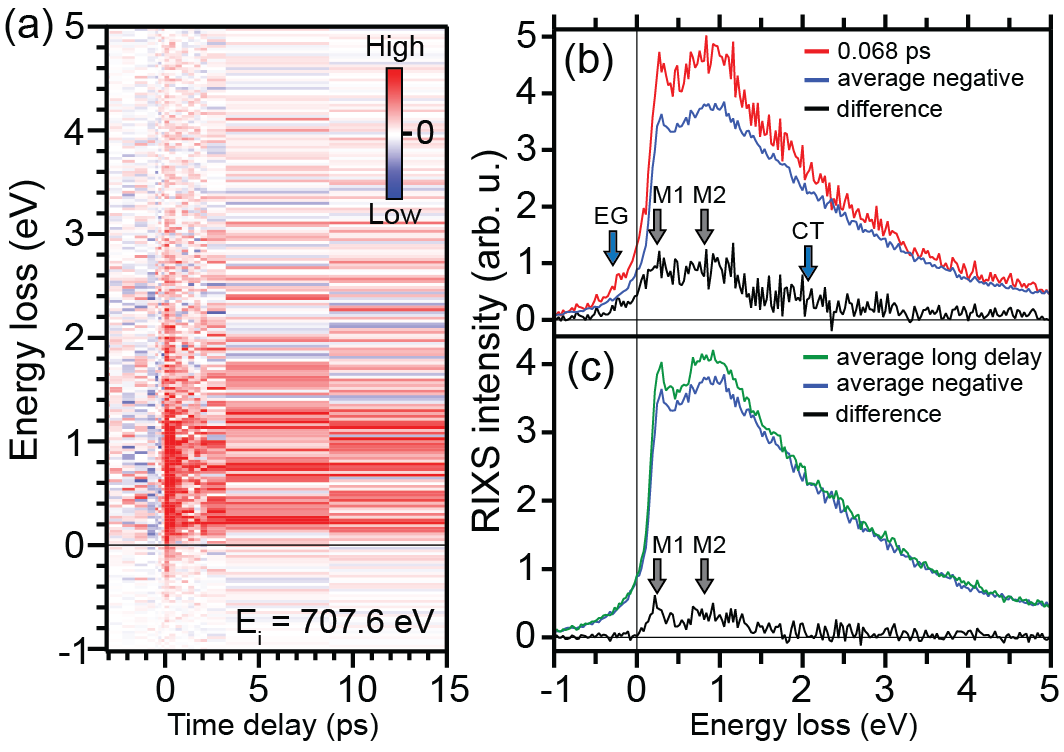}
\caption{Pump-induced intensity changes in RIXS spectra at the XAS pre-edge. (a) Color plot of the difference RIXS energy distribution curves at different delays between the 1~$\mu$m pump pulse and the X-ray probe pulse. The difference RIXS spectra are obtained by subtracting spectrum at each time delay from the average of the negative delay spectra between -3.27~ps and -0.43~ps. (b) Comparison of the pumped RIXS spectrum at a delay of 0.068~ps (red curve) with the average of the negative spectra (blue curve). The difference spectrum is shown as the black curve. EG refers to the energy gain feature and CT refers to the charge transfer excitations. M1 and M2 refer to the low energy molecular orbital excitations (see text). (c) Comparison between the average RIXS spectrum at long delays (green curve) and the average RIXS spectrum at negative delays (blue curve). The average long delay curve is constructed by averaging the RIXS spectra between 13.4~ps and 50.8~ps. The average negative delay curve is obtained the same way as that in panel (b). Energy positions of gray arrows indicating M1 and M2 in (b) and (c) are obtained from fits of the unpumped spectrum. Data in all panels taken at 25~K, incidence angle of 20$\degree$, and optical laser excitation fluence of 11 mJ/cm$^2$.}
\end{figure}

\begin{figure}[t]
\includegraphics[width=0.45\textwidth]{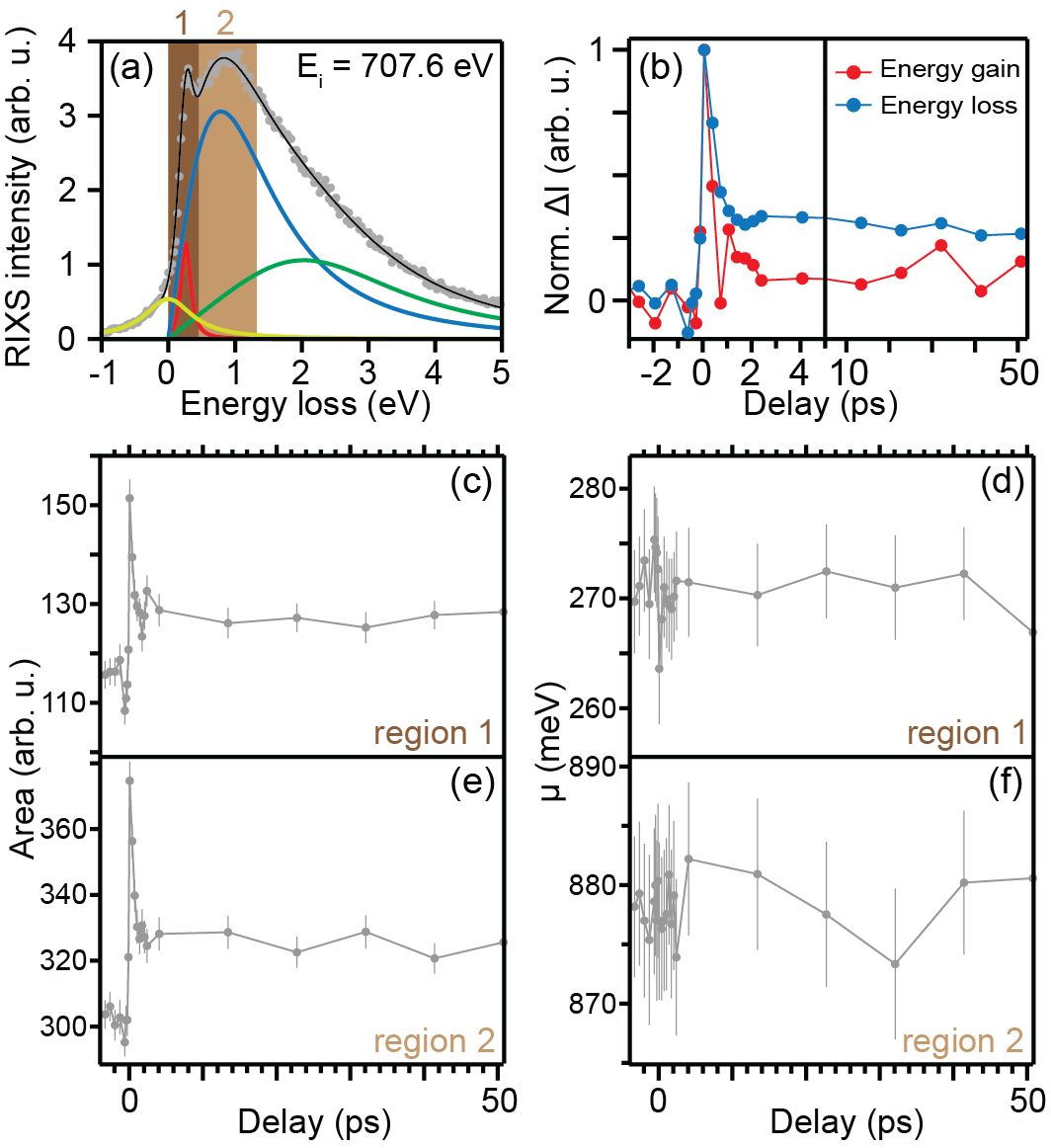}
\caption{Delay dependence of RIXS spectra. (a) Fitting of the unpumped RIXS spectrum (average over the negative delays) to distinguish between different spectral regions. The fit is performed with a Voigt peak for the small quasielastic feature (yellow curve) and anti-Lorentzian peaks for the M1 (red curve), M2 (blue curve), and charge transfer (green curve) features. The overall fit was convolved with the instrument resolution of 175~meV. Brown regions represent regions where the M1 feature is prominent (region 1) and where the M2 feature is prominent (region 2). (b) Delay dependence of the RIXS difference intensity within the energy gain window [-3, 0]~eV (red curve) and energy loss window [0,2]~eV (blue curve). The curves are scaled such that the maximum effect is 1. The difference intensity is obtained by subtracting the spectrum at each time delay by the average of the negative delay spectra. (c, e) Integrated intensity of region 1 (c) and region 2 (e) as a function of delay. (d, f) First moment of the RIXS spectra within region 1 (d) and region 2 (f) as a function of time delay. Error bars in (c-f) represent numerical integration uncertainty obtained from left and right Riemann sums. Data in all panels taken at 25~K, incidence angle of 20$\degree$, and optical laser excitation fluence of 11 mJ/cm$^2$.}
\end{figure}

The delay-dependent XAS intensities at P and M are shown in Fig. 1(d) and 1(e) respectively. For both P and M, we observe a fast, time-resolution-limited onset of the XAS intensity followed first by a fast initial recovery and then a nearly constant response that persists for at least 30 ps. The delay curves are each fitted with a Gaussian rising edge and two exponentially decaying trailing edges. The initial fast recovery time constants are 0.6 $\pm$ 0.1 ps for P and 0.5 $\pm$ 0.3 ps for M. The secondary responses at long time scales have no discernible decay rate within this time frame, as fits yield time constants of larger than 1 ns with large uncertainties. Thus, we characterize the secondary response as a persistent metastable state.

We can obtain an additional dimension of information on the behavior of the response to optical excitation by examining the RIXS energy loss spectrum. Fig. 2(a) shows a color plot of the difference RIXS spectrum as a function of delay between the pump pulse and the X-ray probe pulse at point P. We observe a general strong intensity increase at nearly all energies immediately after arrival of the pump pulse, including a clear feature on the energy gain side. At long delay times, the intensity increase partially recovers and the energy gain feature is not observed anymore. These behaviors are captured in the energy distribution curves shown in panels (b) and (c). At a delay of 0.068 ps, the pumped RIXS spectrum shows an approximately 30\% increase of intensities at the M1 and M2 energies, as well as more subtle increases on the energy gain side and at high energy losses above 1.5~eV.

Fig. 2(c) shows the comparison of the long delay time RIXS spectrum with the negative delay RIXS spectrum. Here, the long delay time RIXS spectrum is constructed by averaging the spectra between 13.4~ps and 50.8~ps delay. This is justified by both the trXAS response shown in Fig. 1(d) and the delay-dependent trRIXS response shown later, where the excited state does not display any significant time dependence within this period. At long delay times, the  transient intensity increase in the energy gain region and the high energy loss regions have nearly fully recovered. However, the intensity increases associated with the M1 and M2 features are still persistent despite a partial recovery.

We can gain additional insights into the dynamics of the photo-excited state by quantitatively examining the RIXS curves as a function of time delay. Fig. 3(a) shows the details of the decomposition of the average negative delay RIXS spectrum through peak fitting guided by the previous RIXS study with higher energy resolution~\cite{kxu1}. The RIXS spectrum fit consists of a small quasi-elastic peak (yellow curve), M1 peak (red curve), M2 peak (blue curve), and high energy charge transfer features (green curve). The small quasi-elastic peak likely mainly arises from the use of two collecting mirrors, which integrates photon counts over about 5$\degree$ of scattering angle. This significantly increases the outgoing photon detection efficiency but introduces elastic contamination from scattering angles that deviate slightly from the 90$\degree$ condition. The quasielastic peak is broader than the 175~meV resolution obtained on the carbon tape, suggesting that additional components such as phonons are present.

The RIXS spectrum fit in Fig. 3(a) gives a good qualitative sense of consistency with previous high-energy-resolution measurements~\cite{kxu1}. Here, we opted to investigate the time-delay response of the RIXS spectrum by examining the intensities within certain windows guided by the fit in panel (a). This is a more robust analysis because the broad charge transfer tail and the higher noise level of individual RIXS spectra result in instabilities during direct fitting. Fig. 3(b) shows a comparison of the integrated difference RIXS intensity on the energy gain side (red curve) and on the energy loss side (blue curve). The energy gain intensity returns to almost the negative delay values after about 2~ps. On the other hand, the difference RIXS intensity recovers by about 70\% by 1 to 2~ps after excitation. However, the remaining 30\% of increased intensity remains persistent for at least 50 ps.

We further examine the M1 and M2 features by calculating the integrated intensities and spectral moments as a function of time delay. Fig. 3(c,e) show that the integrated intensities within windows 1 and 2 (shaded regions in Fig. 3(a)) exhibit the two component recovery behavior. On the other hand, Figs. 3(d,f) show that the first moments of these regions do not display any discernible change within an approximately 5~meV range as a function of time delay, suggesting that the peak position is not changing significantly.

In contrast to the substantial transient RIXS signal at P, we could not resolve a clear transient RIXS signal at M~\cite{supplemental}. We attribute this to two main reasons: the first is that the transient XAS difference signal at M is more spread out in incident energy than at P, resulting in a lower transient signal at a given incident energy near M compared to the maximum value at P. Secondly, the absolute XAS signal is much higher at M than P, resulting in a smaller percentage change for the transient signal. These effects combine to reduce any transient RIXS signal at M below the current achievable noise level within the limited beam time.

The trXAS and trRIXS results here reveal anomalous dynamics of the molecular orbital excitations in FeSb$_2$. In more conventional insulators with well-defined orbital excitations, above-gap optical excitations typically weaken the orbital excitations associated with localized atomic states~\cite{merzoni1} due to screening and damping. At longer time scales, transient lattice expansion can reduce the orbital overlap and soften the energy of the orbital excitations. Another useful reference is the response of many-body Zhang-Rice singlet (ZRS)~\cite{zhang1} excitations in CuGeO$_3$~\cite{paris1}. Here, the pump pulse melts the short-range antiferromagnetic correlations and suppresses the ZRS excitation intensity dramatically. Finally, we can also reference the effects of equilibrium tuning on the molecular orbital excitations in FeSb$_2$. Both chemical doping and increasing temperature result in reduction or broadening of localized excitations~\cite{kxu1}. In comparison, the observed pump-induced intensity increases of the M1 and M2 molecular orbital excitations are unexpected. 

While it is tempting to think about the intensity change in the XAS and RIXS as simple charge transfer and population shifts induced by the pump, such a picture is not suitable in a many-body correlated state. This is because particle-hole excitations in such a system are necessarily accompanied by many-body reconfigurations. Here we provide a possible qualitative framework for understanding the pump-induced XAS and RIXS intensity change without invoking the single particle population language. In the following we define the valence manifold as the Fock space spanned by the single particle state close to the Fermi energy including both the conduction and valence bands.

The XAS process transfers one core electron into the valence manifold, increasing its electron number to $N+1$. The state $|f;h,N+1\rangle$ is the exact final eigenstate in the 1-core hole sector, with $f$ labeling the final state and $h$ specifying the core hole left behind. Core-hole attraction can bind the added electron and reorganize the other electrons through screening and shake ups, meaning the generation of additional particle–hole excitations. The XAS intensity is given by

\begin{equation}
    I(\omega,t)  = \sum_{i,f} p_i(t)|\langle f;h,N+1|T_h|i,N\rangle |^2 \delta(\hbar\omega - [E_h+E_f^{(h)} - E_i])
\end{equation}

Where $T_h = \sum_a M_{ah} d^\dagger_a$ is the projection of the dipole operator onto the valence manifold with $M_{ah}$ being the polarization-dependent dipole matrix element from core level $h$ to valence orbital $a$. $E_h$ is the core hole creation energy; $E_f^{(h)}$ is the eigenenergy of the N+1 electrons in the 1-core hole sector. The exact amplitudes include Pauli blocking, correlations, and core hole shake ups. Enhancement of this XAS intensity can occur when absorption gained from pump-populated excited states exceeds the loss from depleted initial states. The pump-induced excited states $|i,N\rangle$ can produce a greater $T_h$ matrix element with the core hole final state $|f;h,N+1\rangle$ than the unpumped ground state $|0,N\rangle$. This is because the pump photo-excitation can simulate the particle-hole shake up due to the core hole~\cite{supplemental}. For the same final state, an excited initial state requires less photon energy due to its excitation energy. If this difference is small compared to the core-hole broadening, the extra absorption remains approximately within the original peak. Increased intensity of the excited state therefore need not produce a distinct peak but can be consistent with the slight shift observed in Fig. 1.

RIXS emission fills the intermediate core hole and leaves an exact valence state $|m,N\rangle$, labeled by $m$ and having energy $E_m$. In the short core-hole life time limit, the RIXS intensity is

\begin{equation}
    I_{RIXS}(\Omega,t)  \propto \sum_{i,m} p_i(t)|\langle m|O_{eff}|i\rangle |^2 \delta(\hbar\Omega - [E_m - E_i])
\end{equation}

with $\Omega$ being the energy loss. $O_{eff}$ acts on the N electron valence manifold and is approximated by the following when the core hole lifetime is short

\begin{equation}
    O_{eff} \cong C \sum_{h \in \mathbb{R}} T_h^\dagger \textbf{$P_h$} T_h
\end{equation}

where $C = 1 / (\Delta + i\Gamma/2)$ with $\Delta$ being the incident energy detuning and $\Gamma$ being the core hole broadening. $\textbf{$P_h$} = \sum_f |f; h, N+1\rangle \langle f; h, N+1|$ is the operator that projects into the 1-core hole sector.

For the same reason discussed in the XAS, the pump can prepare an initial state that couples more strongly to the core-hole intermediate states through which a particular RIXS excitation is produced. Rather, the particle-hole excitation created by pumping can play the role of spectators which are added to the RIXS excitations such as M1 and M2. As the result, the excitation energy of M1 and M2 can remain the same with pumping. Similar to the XAS case, $T_h$ acting on the pump-prepared state can induce a larger overlap with the intermediate state, hence leading to enhanced matrix element $T_h^\dagger P_h T_h$.

For negative $\Omega$, corresponding to energy gain features, the outgoing photon gains energy as the sample relaxes to a lower energy state. Energy gain features can arise from population of anti-Stokes scattering of collective excitations as seen previously in trRIXS experiments on paramagnons in cuprates~\cite{jost1}. Alternatively, the pump can transiently establish a different electronic configuration than the ground state~\cite{merzoni1}. In FeSb$_2$, the energy gain feature extends to about 0.5 eV, which is larger than any collective excitations such as phonons in the system, favoring a transient occupation of an excited many-body molecular orbital configuration.

After the pump is turned off, a portion of the excited states optically recombine, leading to an initial fast recovery. At long time scales, additional effects may also be important. For example, phonon bottleneck effects~\cite{Rothwarf1,demsar1} can prevent the decay of low energy excited states due to the narrow gap in FeSb$_2$. We further note that additional effects arising from the transient metallization may also be important, as the photo-excited carriers may lead to absorption enhancement through the Mahan–Nozières–de Dominicis effect~\cite{mahan1, roulet1, nozieres1, nozieres2}. In the FeSb$_2$ system, a complete model must also capture itinerant hybridized bands, the local molecular orbitals with appropriate anisotropies, and the many body interactions. Such a theoretical treatment is difficult and beyond the scope of this letter.

Our observation of unusual molecular orbital photo-enhancement in the XAS and RIXS spectra of FeSb$_2$ demonstrates the importance of many-body interactions in X-ray spectroscopy of optically-excited entangled states. Our qualitative physical picture of the pump-induced enhancement of the XAS and RIXS matrix element is an important step towards understanding how the many-body excited state interacts with the X-ray absorption and emission processes. Further works are needed to establish this mechanism through microscopic calculations. Nonetheless, our observations suggest that optical excitations may be able to perturb the localized excitations in FeSb$_2$ in ways not easily accessible with equilibrium methods. A careful selection of pumping fluences and pump photon energies may enable manipulation of the ground state in different regimes.

\begin{acknowledgments}
The authors thank Brian Moritz for fruitful discussions. The work performed at the University of California, Berkeley and Lawrence Berkeley National Laboratory was funded by the U.S. Department of Energy, Office of Science, Office of Basic Energy Sciences, Materials Sciences and Engineering Division, Contract No. DE-AC02-05-CH11231 within the Quantum Materials Program (KC2202). The work performed at Stanford and SLAC was supported by the U.S. Department of Energy (DOE), Office of Basic Energy Sciences, Division of Materials Sciences and Engineering, Contract No. DE-AC02-76SF00515. This work made use of the Furka beamline at the SwissFEL facility under proposal number 20250924. P.M. acknowledges the support from Stanford Energy Postdoctoral Fellowship and the Precourt Institute for Energy.
\end{acknowledgments}

\hspace{1cm}

\bibliographystyle{apsrev4-2}
\bibliography{main_ref}

@PREAMBLE{
 "\providecommand{\noopsort}[1]{}" 
 # "\providecommand{\singleletter}[1]{#1}%" 
}

@ARTICLE{aeppli1,
   author       = "G. Aeppli and Z. Fisk",
   title        = "Kondo Insulators",
   journal      = "Comments Cond. Mat. Phys.",
   volume       = "16", 
   pages        = "155-170",
   year         = "1992",
}

@article{tsunetsugu1,
  title = {The ground-state phase diagram of the one-dimensional Kondo lattice model},
  author = {Tsunetsugu, Hirokazu and Sigrist, Manfred and Ueda, Kazuo},
  journal = {Rev. Mod. Phys.},
  volume = {69},
  issue = {3},
  pages = {809--864},
  numpages = {0},
  year = {1997},
  month = {Jul},
  publisher = {American Physical Society},
  doi = {10.1103/RevModPhys.69.809},
  url = {https://link.aps.org/doi/10.1103/RevModPhys.69.809}
}

@article{menth1,
  title = {Magnetic and Semiconducting Properties of Sm${\mathrm{B}}_{6}$},
  author = {Menth, A. and Buehler, E. and Geballe, T. H.},
  journal = {Phys. Rev. Lett.},
  volume = {22},
  issue = {7},
  pages = {295--297},
  numpages = {0},
  year = {1969},
  month = {Feb},
  publisher = {American Physical Society},
  doi = {10.1103/PhysRevLett.22.295},
  url = {https://link.aps.org/doi/10.1103/PhysRevLett.22.295}
}

@ARTICLE{luli1,
   author       = "L. Li and K. Sun and C. Kurdak and J. W. Allen",
   title        = "Emergent mystery in the Kondo insulator samarium hexaboride",
   journal      = "Nat. Rev. Phys.",
   volume       = "42", 
   pages        = "463-479",
   doi = {10.1038/s42254-020-0210-8},
   year         = "2020",
}

@misc{coleman1,
      title={Heavy Fermions and the Kondo Lattice: a 21st Century Perspective}, 
      author={Piers Coleman},
      year={2015},
      eprint={1509.05769},
      archivePrefix={arXiv},
      primaryClass={cond-mat.str-el},
      url={https://arxiv.org/abs/1509.05769}, 
}

@misc{kxu1,
      title={Spectroscopic evidence for a molecular orbital Kondo insulator}, 
      author={Ke-Jun Xu and Kuan H. Hsu and Nathan Giles-Donovan and Christopher T. Parzyck and Gi-Hyeok Lee and Wanli Yang and Jun Okamoto and Hsiao-Yu Huang and Di-Jing Huang and Joshua J. Kas and John Vinson and Zhi-Xun Shen and Dung-Hai Lee and Thomas P. Devereaux and Wei-Sheng Lee and Robert J. Birgeneau},
      year={2026},
      eprint={2605.30293},
      archivePrefix={arXiv},
      primaryClass={cond-mat.str-el},
      url={https://arxiv.org/abs/2605.30293}, 
}

@ARTICLE{Jaime1,
  title     = "Closing the spin gap in the Kondo insulator {Ce3Bi4Pt3} at high
               magnetic fields",
  author    = "Jaime, M and Movshovich, R and Stewart, G R and Beyermann, W P
               and Berisso, M G and Hundley, M F and Canfield, P C and Sarrao,
               J L",
  journal   = "Nature",
  volume    =  405,
  number    =  6783,
  pages     = "160--163",
  year      =  2000,
  doi = {10.1038/35012027}
}

@article{cooley1,
  title = {Sm${\mathrm{B}}_{6}$: Kondo Insulator or Exotic Metal?},
  author = {Cooley, J. C. and Aronson, M. C. and Fisk, Z. and Canfield, P. C.},
  journal = {Phys. Rev. Lett.},
  volume = {74},
  issue = {9},
  pages = {1629--1632},
  numpages = {0},
  year = {1995},
  month = {Feb},
  publisher = {American Physical Society},
  doi = {10.1103/PhysRevLett.74.1629},
  url = {https://link.aps.org/doi/10.1103/PhysRevLett.74.1629}
}

@article{gabani1,
author = {Slavomír Gabáni and Matúš Orendáč and Gabriel Pristáš and Emil Gažo and Pavel Diko and Samuel Piovarči and Vladimir Glushkov and Nikolay Sluchanko and Anna Levchenko and Natalya Shitsevalova and Karol Flachbart},
title = {Transport properties of variously doped SmB6},
journal = {Philos. Mag.},
volume = {96},
number = {31},
pages = {3274--3283},
year = {2016},
publisher = {Taylor \& Francis},
doi = {10.1080/14786435.2016.1177225},
}

@article{Petrovic1,
title = {Anisotropy and large magnetoresistance in the narrow-gap semiconductor ${\mathrm{FeSb}}_{2}$},
  author = {Petrovic, C. and Kim, J. W. and Bud'ko, S. L. and Goldman, A. I. and Canfield, P. C. and Choe, W. and Miller, G. J.},
  journal = {Phys. Rev. B},
  volume = {67},
  issue = {15},
  pages = {155205},
  numpages = {4},
  year = {2003},
  month = {Apr},
  publisher = {American Physical Society},
  doi = {10.1103/PhysRevB.67.155205},
}

@article{Sun1,
 author = {Sun, P. and Søndergaard, M. and Sun, Y. and Johnsen, S. and Iversen, B. B. and Steglich, F.},
    title = {Unchanged thermopower enhancement at the semiconductor-metal transition in correlated FeSb2−xTex},
    journal = {Appl. Phys. Lett.},
    volume = {98},
    number = {7},
    pages = {072105},
    year = {2011},
    month = {02},
    issn = {0003-6951},
    doi = {10.1063/1.3556645},
}

@misc{Eaton1,
title={Electrical transport signatures of metallic surface state formation in the strongly-correlated insulator FeSb2}, 
      author={Alexander G. Eaton and Nicholas J. M. Popiel and Ke-Jun Xu and Alexander J. Hickey and Hsu Liu and Monica Ciomaga Hatnean and Geetha Balakrishnan and Gunnar F. Lange and Robert-Jan Slager and Zhi-Xun Shen and Suchitra E. Sebastian},
      year={2024},
      eprint={2403.04550},
      archivePrefix={arXiv},
      primaryClass={cond-mat.str-el},
      url={https://arxiv.org/abs/2403.04550}, 
}

@article{Eo1,
author = {Eo, Yun Suk and Avers, Keenan and Horn, Jarryd A. and Yoon, Hyeok and Saha, Shanta R. and Suarez, Alonso and Fuhrer, Michael S. and Paglione, Johnpierre},
    title = {Extraordinary bulk-insulating behavior in the strongly correlated materials FeSi and FeSb2},
    journal = {Appl. Phys. Lett.},
    volume = {122},
    number = {23},
    pages = {233102},
    year = {2023},
    month = {06},
        issn = {0003-6951},
    doi = {10.1063/5.0148249},
}

@article{kxu2,
author = {Ke-Jun Xu  and Su-Di Chen  and Yu He  and Junfeng He  and Shujie Tang  and Chunjing Jia  and Eric Yue Ma  and Sung-Kwan Mo  and Donghui Lu  and Makoto Hashimoto  and Thomas P. Devereaux  and Zhi-Xun Shen },
title = {Metallic surface states in a correlated d-electron topological Kondo insulator candidate FeSb<sub>2</sub>},
journal = {Proc. Natl. Acad. Sci.},
volume = {117},
number = {27},
pages = {15409-15413},
year = {2020},
doi = {10.1073/pnas.2002361117},
}

@article{Petrovic2,
  title = {Kondo insulator description of spin state transition in $\mathrm{Fe}{\mathrm{Sb}}_{2}$},
  author = {Petrovic, C. and Lee, Y. and Vogt, T. and Lazarov, N. Dj. and Bud'ko, S. L. and Canfield, P. C.},
  journal = {Phys. Rev. B},
  volume = {72},
  issue = {4},
  pages = {045103},
  numpages = {7},
  year = {2005},
  month = {Jul},
  publisher = {American Physical Society},
  doi = {10.1103/PhysRevB.72.045103},
  url = {https://link.aps.org/doi/10.1103/PhysRevB.72.045103}
}

@article{Sun2,
author = {Sun, P. and Søndergaard, M. and Iversen, B.B. and Steglich, F.},
title = {Strong electron correlations in FeSb2},
journal = {Ann. Phys. (Berl.)},
volume = {523},
number = {8-9},
pages = {612-620},
doi = {https://doi.org/10.1002/andp.201100033},
year = {2011},
}

@article{Vinson1,
title = {Bethe-Salpeter equation calculations of core excitation spectra},
  author = {Vinson, J. and Rehr, J. J. and Kas, J. J. and Shirley, E. L.},
  journal = {Phys. Rev. B},
  volume = {83},
  issue = {11},
  pages = {115106},
  numpages = {7},
  year = {2011},
  month = {Mar},
  publisher = {American Physical Society},
  doi = {10.1103/PhysRevB.83.115106},
  url = {https://link.aps.org/doi/10.1103/PhysRevB.83.115106}
}

@article{Vinson2,
author = {Vinson, John},
    title = {Advances in the ocean-3 spectroscopy package},
    journal = {Phys. Chem. Chem. Phys.},
    volume = {24},
    number = {21},
    pages = {12787-12803},
    year = {2022},
    month = {06},
    issn = {1463-9076},
    doi = {10.1039/d2cp01030e},
    url = {https://doi.org/10.1039/d2cp01030e},
}

@ARTICLE{degroot1,
title = {Multiplet effects in X-ray spectroscopy},
journal = {Coord. Chem. Rev.},
volume = {249},
number = {1},
pages = {31-63},
year = {2005},
issn = {0010-8545},
doi = {https://doi.org/10.1016/j.ccr.2004.03.018},
author = {Frank de Groot},
}

@article{Ament1,
title = {Resonant inelastic x-ray scattering studies of elementary excitations},
  author = {Ament, Luuk J. P. and van Veenendaal, Michel and Devereaux, Thomas P. and Hill, John P. and van den Brink, Jeroen},
  journal = {Rev. Mod. Phys.},
  volume = {83},
  issue = {2},
  pages = {705--767},
  numpages = {0},
  year = {2011},
  month = {Jun},
  publisher = {American Physical Society},
  doi = {10.1103/RevModPhys.83.705},
  url = {https://link.aps.org/doi/10.1103/RevModPhys.83.705}
}

@article{Kikoin1,
  author    = {Kikoin, K. A. and Mishchenko, A. S.},
  title     = {Magnetic excitations in intermediate-valence semiconductors with a singlet ground state},
  journal   = {J. Phys.: Condens. Matter},
  volume    = {7},
  number    = {2},
  pages     = {307--325},
  year      = {1995},
  doi       = {10.1088/0953-8984/7/2/008}
}

@inproceedings{razzoli1,
author = {E. Razzoli and H. Ueda and E. Paris and B. Liu and E. Skoropata and L. Patthey},
title = {{Ultrafast dynamics in quantum matter at SwissFEL: capabilities of Furka endstation}},
volume = {12581},
booktitle = {X-Ray Free-Electron Lasers: Advances in Source Development and Instrumentation VI},
organization = {International Society for Optics and Photonics},
publisher = {SPIE},
pages = {125810D},
year = {2023},
doi = {10.1117/12.2668886},
URL = {https://doi.org/10.1117/12.2668886}
}

@article{Homes1,
  author    = {Homes, C. C. and Du, Q. and Petrovic, C. and Brito, W. H. and Choi, S. and Kotliar, G.},
  title     = {Unusual electronic and vibrational properties in the colossal thermopower material {FeSb$_2$}},
  journal   = {Sci. Rep.},
  volume    = {8},
  number    = {1},
  pages     = {11692},
  year      = {2018},
  doi       = {10.1038/s41598-018-29909-2}
}

@misc{merzoni1,
      title={Photo-generated charge-transfer excitons in NiO revealed by ultrafast time-resolved resonant inelastic x-ray scattering}, 
      author={Giacomo Merzoni and Leonardo Martinelli and Sergii Parchenko and Sophia F. R. TenHuisen and Vasily Lebedev and Luigi Adriano and Robert Carley and Natalia Gerasimova and Laurent Mercadier and Martin Teichmann and Benjamin E. van Kuiken and Zhong Yin and Amina Alic and Denitsa R. Baykusheva and Sorin G. Chiuzbaian and Stefano Dal Conte and Oleg Dogadov and Alexander Föhlisch and Maurits W. Haverkort and Maximilian Kusch and Tim Laarmann and Wei Sheng Lee and Marco Moretti Sala and Ying Ying Peng and Qing Zheng Qiu and Thorsten Schmitt and Sreeju Sreekantan Nair Lalithambika and Simone Techert and Giulio Cerullo and Michael Först and Matteo Mitrano and Mark P. M. Dean and Justine Schlappa and Andreas Scherz and Giacomo Ghiringhelli},
      year={2025},
      eprint={2504.16653},
      archivePrefix={arXiv},
      primaryClass={cond-mat.str-el},
      url={https://arxiv.org/abs/2504.16653}, 
}

@ARTICLE{Paris1,
  title     = {Probing the interplay between lattice dynamics and short-range magnetic correlations in CuGeO3 with femtosecond RIXS},
  author    = {Paris, E and Nicholson, C W and Johnston, S and Tseng, Y and
               Rumo, M and Coslovich, G and Zohar, S and Lin, M F and Strocov,
               V N and Saint-Martin, R and Revcolevschi, A and Kemper, A and
               Schlotter, W and Dakovski, G L and Monney, C and Schmitt, T},
  journal   = {npj Quantum Mater.},
  volume    =  {6},
  number    =  {1},
  pages     = {51},
  year      =  {2021},
  doi       = {10.1038/s41535-021-00350-5}
}

@article{zhang1,
  title = {Effective Hamiltonian for the superconducting Cu oxides},
  author = {Zhang, F. C. and Rice, T. M.},
  journal = {Phys. Rev. B},
  volume = {37},
  issue = {7},
  pages = {3759(R)--3761(R)},
  numpages = {0},
  year = {1988},
  month = {Mar},
  publisher = {American Physical Society},
  doi = {10.1103/PhysRevB.37.3759},
  url = {https://link.aps.org/doi/10.1103/PhysRevB.37.3759}
}

@article{Rothwarf1,
  title = {Measurement of Recombination Lifetimes in Superconductors},
  author = {Rothwarf, Allen and Taylor, B. N.},
  journal = {Phys. Rev. Lett.},
  volume = {19},
  issue = {1},
  pages = {27--30},
  numpages = {0},
  year = {1967},
  month = {Jul},
  publisher = {American Physical Society},
  doi = {10.1103/PhysRevLett.19.27},
  url = {https://link.aps.org/doi/10.1103/PhysRevLett.19.27}
}

@article{demsar1,
  title = {Photoexcited Electron Dynamics in Kondo Insulators and Heavy Fermions},
  author = {Demsar, Jure and Thorsm\o{}lle, Verner K. and Sarrao, John L. and Taylor, Antoinette J.},
  journal = {Phys. Rev. Lett.},
  volume = {96},
  issue = {3},
  pages = {037401},
  year = {2006},
  doi = {10.1103/PhysRevLett.96.037401},
  url = {https://link.aps.org/doi/10.1103/PhysRevLett.96.037401}
}

@article{jost1,
  title = {Collective Magnetic Excitations in a Photoexcited Electron-Doped Cuprate Superconductor},
  author = {Jost, Daniel and Li, Jiarui and Hales, Jordyn and Sobota, Jonathan and Merzoni, Giacomo and Martinelli, Leonardo and Ding, Shuhan and Xu, Ke-Jun and Schlappa, Justine and Scherz, Andreas and Carley, Robert and Van Kuiken, Benjamin E. and Asmara, Teguh C. and Hoang, Le Phuong and Mercadier, Laurent and Parchenko, Sergii and Teichmann, Martin and Kirchmann, Patrick S. and Ghiringhelli, Giacomo and Moritz, Brian and Shen, Zhi-Xun and Devereaux, Thomas P. and Wang, Yao and Lee, Wei-Sheng},
  journal = {Phys. Rev. Lett.},
  volume = {137},
  issue = {9},
  pages = {096502},
  numpages = {7},
  year = {2026},
  month = {Aug},
  publisher = {American Physical Society},
  doi = {10.1103/2h9x-8tjk},
  url = {https://link.aps.org/doi/10.1103/2h9x-8tjk}
}

@article{mahan1,
  title = {Excitons in Metals: Infinite Hole Mass},
  author = {Mahan, G. D.},
  journal = {Phys. Rev.},
  volume = {163},
  issue = {3},
  pages = {612--617},
  numpages = {0},
  year = {1967},
  month = {Nov},
  publisher = {American Physical Society},
  doi = {10.1103/PhysRev.163.612},
  url = {https://link.aps.org/doi/10.1103/PhysRev.163.612}
}

@article{roulet1,
  title = {Singularities in the X-Ray Absorption and Emission of Metals. I. First-Order Parquet Calculation},
  author = {Roulet, B. and Gavoret, J. and Nozi{\`e}res, P.},
  journal = {Phys. Rev.},
  volume = {178},
  issue = {3},
  pages = {1072--1083},
  numpages = {0},
  year = {1969},
  month = {Feb},
  publisher = {American Physical Society},
  doi = {10.1103/PhysRev.178.1072},
  url = {https://link.aps.org/doi/10.1103/PhysRev.178.1072}
}

@article{nozieres1,
  title = {Singularities in the X-Ray Absorption and Emission of Metals. II. Self-Consistent Treatment of Divergences},
  author = {Nozi{\`e}res, P. and Gavoret, J. and Roulet, B.},
  journal = {Phys. Rev.},
  volume = {178},
  issue = {3},
  pages = {1084--1096},
  numpages = {0},
  year = {1969},
  month = {Feb},
  publisher = {American Physical Society},
  doi = {10.1103/PhysRev.178.1084},
  url = {https://link.aps.org/doi/10.1103/PhysRev.178.1084}
}

@article{nozieres2,
  title = {Singularities in the X-Ray Absorption and Emission of Metals. III. One-Body Theory Exact Solution},
  author = {Nozi{\`e}res, P. and de Dominicis, C. T.},
  journal = {Phys. Rev.},
  volume = {178},
  issue = {3},
  pages = {1097--1107},
  numpages = {0},
  year = {1969},
  month = {Feb},
  publisher = {American Physical Society},
  doi = {10.1103/PhysRev.178.1097},
  url = {https://link.aps.org/doi/10.1103/PhysRev.178.1097}
}

@article{mignot2,
  title = {Evidence for Short-Range Antiferromagnetic Fluctuations in Kondo-Insulating ${\mathrm{YbB}}_{12}$},
  author = {Mignot, J.-M. and Alekseev, P. A. and Nemkovski, K. S. and Regnault, L.-P. and Iga, F. and Takabatake, T.},
  journal = {Phys. Rev. Lett.},
  volume = {94},
  issue = {24},
  pages = {247204},
  numpages = {4},
  year = {2005},
  month = {Jun},
  publisher = {American Physical Society},
  doi = {10.1103/PhysRevLett.94.247204},
  url = {https://link.aps.org/doi/10.1103/PhysRevLett.94.247204}
}

@ARTICLE{Rahn1,
  title     = "Kondo quasiparticle dynamics observed by resonant inelastic
               x-ray scattering",
  author    = "Rahn, M C and Kummer, K and Hariki, A and Ahn, K-H and Kune{\v
               s}, J and Amorese, A and Denlinger, J D and Lu, D-H and
               Hashimoto, M and Rienks, E and Valvidares, M and Haslbeck, F and
               Byler, D D and McClellan, K J and Bauer, E D and Zhu, J X and
               Booth, C H and Christianson, A D and Lawrence, J M and Ronning,
               F and Janoschek, M",
  journal   = "Nat. Commun.",
  volume    =  13,
  number    =  1,
  pages     = "6129",
  year      =  2022,
doi = "10.1038/s41467-022-33468-6"
}

@misc{supplemental,
  note = "See Supplemental Materials at (link to be inserted by publisher)."
}
\raggedend
\end{document}